\documentclass[amsmath,amssymb,aps,pra,twocolumn,showpacs,longbibliography]{revtex4-1}
\PassOptionsToPackage{breaklinks}{hyperref}
\AtBeginDocument{\RequirePackage{hyperref}}

\usepackage{verbatim}

\usepackage[utf8]{inputenc}
\newcommand{\ds}{\displaystyle}
\newcommand{\vp}{\varphi}
\DeclareMathOperator{\tr}{tr}
\newcommand{\vt}{\vartheta}

\newcommand{\bgeq}{\begin{equation}}
\newcommand{\edeq}{\end{equation}}
\DeclareMathOperator{\sinc}{sinc}
\usepackage{amsmath}
\begin{document}
\title{Wigner functions for angle and orbital
angular momentum: Operators and dynamics}
\author{H. A. Kastrup}
\email{hans.kastrup@desy.de}
\affiliation{DESY, Theory Group, Notkestrasse 85, D-22607 Hamburg, Germany}
\begin{abstract}
  Recently a paper on the contruction of  consistent Wigner functions for cylindrical phase spaces $\mathbb{S}^1 \times \mathbb{R}$, i.e.\
for the canonical pair angle and orbital angular momentum, was presented,  main properties of those functions derived, discussed and their usefulness illustrated
by examples. The present paper is a continuation which compares properties of the new Wigner functions for cylindrical phase spaces
with those of the well-known Wigner functions on planar ones in more detail. Furthermore, the mutual (Weyl) correspondence between HIlbert space
operators and  their phase space functions  is discussed. The  $\star$ product formalism is shown to be completely implementable.
 In addition  basic dynamical laws for  Wigner and Moyal functions
 are derived as generalized Liouville and energy equations. They are very similar to those of the planar
case, but also show characteristic differences.
\end{abstract}
\maketitle
\section{Introduction}

In a recent paper \cite{ka2}  mathematically and physically consistent Wigner functions for cylindrical phase spaces 
$\mathbb{S}^1 \times \mathbb{R}$ were proposed, based to a large extent on group theoretical considerations: Replacing the angle $\theta$ - which
characterizes  the $\mathbb{S}^1$-part - by the pair $(\cos\theta, \sin\theta)$  allows for a consistent quantization
in terms of unitary representations of the Euclidean group $E(2)$ of the plane. This group comes into play because the Poisson
brackets of $\cos\theta,\, \sin\theta$ and the angular momentum $p$ obey the Lie algebra of $E(2)$ so that the corresponding quantum 
mechanical self-adjoint operators (observables) $C$, $S$ and $L$ become the generators of unitary representations of $E(2)$ \cite{ka}.

In view of the many applications of Wigner and Moyal functions for planar phase spaces, see, e.g., 
Refs.\ \cite{wig,groo,car,wig1,coh2,lee,leo,schl,roe,leo2,aga,coh},
 a similarly well-founded theoretical framework
for cylindrical ones may open paths to new applications in physics, mathematics, informatics and technologies.
A few simple typical examples are discussed in Ref.\ \cite{ka2}. In  physics, additional applications to fractional
orbital angular momenta \cite{ka} or pendulum-type systems might be of interest.

Considering the structural similarities between the well known Wigner-Moyal functions for planar phase spaces and those for cylindrical phase spaces proposed
in Ref.\ \cite{ka2}, it is a strong challenge to apply the many methods and tools developed for the former now to the cylindrical case.

The Wigner functions $V_{\psi}(\theta,p) =(\psi,V(\theta,p)\psi)$, the more general Moyal functions $V_{\psi_2\psi_1}(\theta,p) =
(\psi_2,V(\theta,p)\psi_1)$ and the associated Wigner Hermitean operator/matrix $V(\theta,p)=(V_{mn}(\theta,p))$ introduced in Ref.\ \cite{ka2} are constructed  
by a kind of ordered group-averaging within a Hilbert space
$L^2(S^1,d\vp/2\pi)$ with the scalar product
\begin{equation}
  \label{eq:1}
  (\psi_2,\psi_1)_{\mathbb{S}^1} =
  \int_{-\pi}^{\pi}\frac{d\vp}{2\pi}\psi_2^{\ast}(\vp)\psi_1(\vp)\,,
\end{equation}
and a basis
\begin{equation}
  \label{eq:2}
  e_{n}(\vp)=e^{ i n\vp},~~(e_{m},
  e_{n})_{\mathbb{S}^1  }=\delta_{mn},\,m,n \in \mathbb{Z}\,,
\end{equation}
where $\delta_{mn}$ is the usual Kronecker symbol. (Generally one has a basis $e_{n,\delta}(\vp) = \exp[i(n+ \delta)],\, \delta \in [0, 1)$,
where $\delta$ characterizes a covering group of the rotation group $SO(2)$ \cite{ka,ka2}. Here we consider only the case $\delta =0$.)
 
We then have 
\begin{equation}
  \label{eq:125}
 V_{mn}(\theta,p) = (e_m,V(\theta,p)e_n)_{\mathbb{S}^1} 
\end{equation}
and 
\begin{equation} \psi(\vp) = \sum_{n \in \mathbb{Z}}c_n\,e_{n}(\vp)\,,~c_n =
 (e_n,\psi)_{\mathbb{S}^1}. \label{eq:3} \end{equation}
The matrix elements $V_{mn}(\theta,p)$ have the explicit form \cite{ka2}
\begin{eqnarray}  
  V_{m n}(\theta,p) =  
 && \frac{1}{2\pi}\,e^{ i(n-m)\theta}\int_{-\pi}^{\pi}\frac{d\vt}{2\pi}\,
  e^{ i[(n+m)/2
    -p/\hbar]\vt} \label{eq:4}\\ &&  =
  \frac{1}{2\pi}\,e^{ i(n-m)\theta}\,\sinc\pi[p/\hbar-(m+n)/2],~~~~  \label{eq:5} \\
&& \sinc x \equiv \frac{\sin x}{x}, \nonumber \end{eqnarray} and yield the Wigner function - a bilinear form - for a wave function $\psi(\vp)$ \begin{eqnarray}
V_{\psi}(\theta,p) =&& \sum_{m,n \in\mathbb{Z}}c_m^{\ast}V_{mn}(\theta,p)c_n \label{eq:17} \\
=&& \frac{1}{2\pi}\int_{-\pi}^{\pi}\frac{d\vt}{2\pi}\,e^{-ip\vt/\hbar}\psi^{\ast}(\theta-\vt/2)\psi(\theta+\vt/2).~~~~ \nonumber
\end{eqnarray}
Many properties of $V_{mn}(\theta,p)$ are discussed in Ref.\ \cite{ka2}. Those which are
essential for the present paper will be listed below.

In Ref.\ \cite{ka2} it was pointed out - without going into details - that many properties
associated with the matrix \eqref{eq:4} and the Wigner function \eqref{eq:17} correspond to structurally equivalent ones
for the usual Wigner functions on the plane $\mathbb{R}^2$: \begin{equation} \label{eq:6}
W_{\phi}(q,p) = \frac{1}{2\pi\hbar}\int_{-\infty}^{\infty}d\xi\,e^{-ip\xi/\hbar}\phi^{\ast}(q-\xi/2)\phi(q+\xi/2).
\end{equation}
(The letter ``$V$'' introduced in Ref.\ \cite{ka2} for the ``cylindrical'' Wigner functions  stands for ``V''ortex or ``V''ariant of the ``planar'' ``$W$''.)

Superficially the expressions \eqref{eq:17} and \eqref{eq:6} look very similar: The integral in Eq.\ \eqref{eq:17} appears as a restriction from the one over
 $\mathbb{R}$
in Eq.\ \eqref{eq:6} to one over the finite interval $[-\pi,\pi]$. The implications are, however, quite non-trivial: Whereas there exists a well-defined position
operator $Q$ in the quantum version of the coordinate $q$, no such operator does exist for the angle $\theta$ \cite{ka3}, and the angle has to be replaced by the 
equivalent pair $(\cos\theta, \sin\theta)$ before quantization. Furthermore,  the translations $q \to q + a$ or $\xi \to \xi + a$ form an invariance group of
 the integral \eqref{eq:6}.
The corresponding transformations for the integral \eqref{eq:17} are rotations from the group $SO(2)$: $\theta \to \theta + \alpha$ or $\vt \to \vt + \alpha$.
 For the associated
quantum mechanics this means that the angular momentum operator $L $, the canonically conjugate variable of the angle, has a discrete spectrum, whereas  momentum
 operator $P$ and position operator $Q$ of the phase plane generally have a continuous one. In addition, the quantum mechanics (representation theory) of the
  group $SO(2)$ has subtleties the translations 
do not have:
 $SO(2)$ is infinitely connected - you can wrap a string around a cylinder an infinite number of times -,  which implies an infinite number of covering groups
and, therefore,  allowing for fractional angular momenta \cite{ka} and so for additional physical effects! 
 
The close correspondence between quite a number of relations of the two types of Wigner functions will be made more explicit below in Sec. II.
Furthermore, as the discussions of relations between operators in Hilbert space and their corresponding functions on phase space
are still incomplete in Ref.\ \cite{ka2}, they will be discussed  in more detail in Sec. III. Sec. IV contains an explicit
 analysis of the dynamics - time evolution and energy equations - of the Wigner-Moyal functions for the cylindrical case.
\section{Comparisons of Wigner functions\\ for planar and  cylindrical\\ phase spaces}
In this section well-known results for the planar case, see, e.g., Refs.\ \cite{groo,wig1,fol,gos,zac,case,zac2}, are compared to those established in Ref.\ \cite{ka2}.
Additional {\it new} results for the cylindrical case are discussed in Secs. III and IV and compared to corresponding properties of the planar one.

In the cylindrical case operators $A$ are generally and conveniently represented by infinite matrices: 
\begin{equation}
  \label{eq:126}
  A = (A_{mn} = (e_m,Ae_n)_{\mathbb{S}^1}),~m,n \in \mathbb{Z}.
\end{equation}
In the following, wave functions in the Hilbert space on $\mathbb{R}$ are denoted by $\phi(x)$, those on $\mathbb{S}^1$ by $\psi(\vp)$.

 Another remark concerns the
different incorporations of Planck's constant $\hbar$: Comparing the {\it integrals} \eqref{eq:6} and \eqref{eq:7} shows that the former is dimensionless. The prefactor
 $1/(2\pi\hbar)$ serves the following purpose: The integration measure on a planar phase space, $dq\,dp$,  has the dimension of an action.  
It is usually made dimensionless by dividing it by $2\pi \hbar = h$.
 However, only the
product $dq\,dp$ itself has the dimension of an action. So one cannot associate that denominator with $dq$ or $dp$ alone. But there are situations where one would like to
integrate over $q$ or $p$ separately. In order to avoid the problem mentioned one inserts the factor $1/(2\pi\hbar$ into the integrand \eqref{eq:6}. In this way
we get the (dimensionless) normalization
\begin{equation}
  \label{eq:131}
  \int_{\mathbb{R}^2} dq\,dp\,W_{\phi}(q,p) = 1.
\end{equation}

 On the other hand,
in the case of a cylindrical phase space with measure $d\theta\,dp$ the angular momentum $p$ alone has the dimension of an action. Thus, one makes $dp$ dimensionless
by dividing it by $\hbar$ and associates the remaining factor $1/(2\pi)$ with the dimensionless $d\theta$. For convenience we write
\begin{equation}
  \label{eq:132}
  d\bar{p} \equiv d(p/\hbar)
\end{equation}
 and absorb the factor $1/(2\pi)$ into the integrand \eqref{eq:17}. We then have the normalization
 \begin{equation}
   \label{eq:133}
   \int_{-\pi}^{\pi}d\theta\,\int_{-\infty}^{\infty}d\bar{p}\,V_{\psi}(\theta,p) = 1.
 \end{equation}
In order to keep the appearance of $\hbar$ otherwise explicit in the following, we  do not in general replace $p/\hbar$ in functions or integrands by $\bar{p}$.
This serves the discussions of classical limits.

 \subsection{Wigner functions proper} 
\subsubsection{Expectation values of operators and \\ operator--related phase-space functions}
The basic postulate for the concept of Wigner functions - here for the planar case - is to express the expectation value $\langle A\rangle_{\phi}$ of an
operator $A$ with respect to a state $\phi$ in a Hilbert space with scalar product
\begin{equation}  \label{eq:7}
(\phi_2,\phi_1)_{\mathbb{R}} = \int_{-\infty}^{\infty}dx\,\phi_2^{\ast}(x)\phi_1(x)
\end{equation}
as an integral over phase space with density $W_{\phi}(q,p)$: 
\begin{equation}
  \label{eq:8}
  \langle A \rangle_{\phi} = (\phi,A\phi)_{\mathbb{R}} = \int_{\mathbb{R}^2}dq\,dp\,
 \tilde{A}(q,p)\,W_{\phi}(q,p),
\end{equation}
where, in Dirac's notation,
\begin{equation}
  \label{eq:9}
\tilde{A}(q,p) = \int_{-\infty}^{\infty}d\xi e^{-ip\xi/\hbar}\langle q+\xi/2|A|q-\xi/2\rangle  
\end{equation}
is the phase space function associated with the Hilbert space operator $A$.

The expression \eqref{eq:9} for $\tilde{A}(q,p)$ may be rewritten in a way which resembles the
corresponding cylindrical one very closely: 
Introducing the operator\cite{leaf1,leaf2,groo}
\begin{equation}
  \label{eq:10}
  \Delta(q,p) = \int_{-\infty}^{\infty}d\xi e^{-ip\xi/\hbar}|q-\xi/2\rangle\langle q+\xi/2|,
\end{equation}
the function $\tilde{A}(q,p)$ can be written as
\begin{equation}
  \label{eq:11}
  \tilde{A}(q,p) = \tr[A\cdot\Delta(q,p)],
\end{equation}
where 
\begin{equation}
  \label{eq:12}
  \tr(B) =\int_{-\infty}^{\infty}dx\,\langle x|B|x\rangle.
\end{equation}
Expectation values of operators with respect to a density operator $\rho$ will be treated below
when traces of operators and their products are discussed.

If $A$ is an operator in a Hilbert space with scalar product \eqref{eq:1} then -
according to Ref.\ \cite{ka2} - its expectation value with respect to a wave function
$\psi(\vp)$ is given by an associated integral over the cylindrical phase space:
\begin{equation}
  \label{eq:13}
  \langle A \rangle_{\psi} = (\psi,A\psi)_{\mathbb{S}^1} = \int_{-\infty}^{\infty}d\bar{p}\, \int_{-\pi}^{\pi}d\theta\, \tilde{A}(\theta,p)\,V_{\psi}(\theta,p),
\end{equation}
where now
\begin{equation}
  \label{eq:14}
  \tilde{A}(\theta,p) = 2\pi \tr[A\cdot V(\theta,p)].
\end{equation}
As $V(\theta + 2\pi,p) = V(\theta,p)$ (see Eq.\ \eqref{eq:4}), the function $\tilde{A}(\theta,p)$ has the same periodicity.

For, e.g., the elementary operators
\begin{equation}
  \label{eq:15}
  C=\cos\vp,~~S=\sin\vp,~~L = \frac{\hbar}{i}\partial_{\vp}
\end{equation}one has - according to Eq.\ \eqref{eq:126} - the matrix elements
\begin{eqnarray}
  \label{eq:127}
C_{mn}& =& \frac{1}{2}(\delta_{m(n+1)} + \delta_{m(n-1}),\\ ~S_{mn}& =& \frac{1}{2i}(\delta_{m(n+1)} - \delta_{m(n-1}),\nonumber \\ ~L_{mn}& =&\hbar\, m\,\delta_{mn}.\nonumber
\end{eqnarray}
Using the relation
\begin{equation}
  \label{eq:118}
  \frac{1}{2\pi}\sum_{n \in \mathbb{Z}}e^{\ds i n\beta} = 
  \delta(\beta)  \mbox{ for } \beta \in [-\pi,\,+\pi], 
\end{equation}
one obtains, according to Eq.\ \eqref{eq:14},
\begin{eqnarray}
  \label{eq:16}
\tilde{C}(\theta,p) &=&2\pi  \tr[C\cdot V(\theta,p)] = \cos\theta, \\ \tilde{S}(\theta,p)& =& 2\pi \tr[S\cdot V(\theta,p)] = \sin\theta,  \\
\tilde{L}(\theta,p)& =& 2\pi \tr[L\cdot V(\theta,p)] = p.
\end{eqnarray}
As a real $2\pi$-periodic potential $U(\vp)$ of a Hamilton operator
\begin{equation}
  \label{eq:21}
  H(\vp,L) = \gamma\,L^2 + U(\vp)
\end{equation}
can be expanded in terms of the multiplication operators
\begin{equation}
  \label{eq:130}
 C^{(k)} = \cos k\vp,~ S^{(k)} = \sin k\vp,~~ k \in \mathbb{Z},
\end{equation}
 the following
relations are useful, too:
\begin{eqnarray}
  \label{eq:22}
 \tilde{C^{(k)}}(\theta,p)& =& 2\pi \tr[V(\theta,p)\cdot C^{(k)}] =  \cos k\theta, \\ 
 \tilde{S^{(k)}}(\theta,p)& =& 2\pi \tr[V(\theta,p)\cdot S^{(k)}] =  \sin k\theta, \\
 \tilde{U}(\theta,p)& =& 2\pi \tr[V(\theta,p)\cdot U] =  U(\theta) \\
 \tilde{L^2}(\theta,p)& =& 2\pi \tr[V(\theta,p)\cdot L^2] =  p^2,\label{eq:136} \\
 \tilde{H}(\theta,p)& =& 2\pi \tr[V(\theta,p)\cdot H] =  H(\theta,p).\label{eq:151}
\end{eqnarray}
In the same manner one gets for the  ``mixed'' products $L\cdot C$ and $C\cdot L$:
\begin{eqnarray}
  \label{eq:128}
 \widetilde{LC}(\theta,p) &=& 2\pi \tr[L\cdot C\cdot V(\theta,p) = p\cos\theta + i\frac{\hbar}{2}\sin\theta,~~~~ \\
\widetilde{CL}(\theta,p) &=& 2\pi\tr[C\cdot L\cdot V(\theta,p) = p\cos\theta - i\frac{\hbar}{2}\sin\theta,~~~~ \label{eq:149}
\end{eqnarray}
so that
\begin{eqnarray}
  \label{eq:129}
\widetilde{[L,C]}(\theta,p) &=& 2\pi  \tr\{[L,C]\cdot V(\theta,p)\} = i \hbar\sin\theta,\\
\widetilde{\{L,C\}}(\theta,p) &=& 2\pi  \tr[\{L,C\}\cdot V(\theta,p)] = 2p\cos\theta.\label{eq:134}
\end{eqnarray}
Eq.\ \eqref{eq:129} corresponds to the Lie algebra relation $[L,C] =\hbar i S$ of the operators \eqref{eq:15} or the Poisson
bracket $\{p,\cos\theta\} = \sin \theta$ \cite{ka,ka2}. The occurrence of $\hbar$ in Eq.\ \eqref{eq:129} shows its quantum mechanical character. In
the limit $\hbar \to 0$ the operators $L$ and $C$ commute, i.e., they become ``classical''. 

The anticommutator function \eqref{eq:134} does not contain any $\hbar$ and is purely ``classical''.

The general case of phase space functions for  products of operators will be discussed in Sec.\ III below.

The phase space function $\tilde{V}(\theta,p)$ of the operator/matrix $V(\theta_1,p_1)$ itself is
\begin{gather}
  \label{eq:63}
  \tilde{V}(\theta,p) = 2\pi\tr[V(\theta,p)\cdot V(\theta_1,p_1)]= \\ = \delta(\theta-\theta_1)\sinc[\pi(p-p_1)/\hbar].\nonumber
\end{gather}
The corresponding relation for the planar space operator \eqref{eq:10} is, using Eq.\ \eqref{eq:11}:
\begin{equation}
  \label{eq:64}
  \tilde{\Delta}(q,p) = \tr[\Delta(q,p)\cdot \Delta(q_1, p_1)] = 2\pi \hbar\, \delta(q-q_1)\,\delta(p-p_1).
\end{equation}
The inverse problem, namely constructing the Hilbert space operator $A$ from a given phase space function $\tilde{A}(\theta,p)$
will be addressed in Sec.\ III.
\subsubsection{Marginal distributions and transition probabilities}
As the Wigner functions \eqref{eq:6} and \eqref{eq:17} can be negative on certain subsets
of their respective phase spaces, they are no proper probability distributions. That $W_{\phi}(q,p)$
may be negative can be seen from the relation
\begin{equation}
  \label{eq:18}
  W_{\phi}(q=0,p=0)=\frac{1}{2\pi\hbar}\int_{-\infty}^{\infty}d\xi\, \phi^{\ast}(-\xi/2)\phi(\xi/2).
\end{equation}
So, if $\phi(x)$ is an odd function, the right hand side of the last equation is negative.
This is, e.g., so for the odd $n$ eigenfunctions of the harmonic oscillator \cite{schl,aga}.
Similar arguments apply to the Wigner function \eqref{eq:17}. In addition it follows from Eq.\ \eqref{eq:5}
that the Wigner function $V_{mm}(\theta,p)$ of the basic function $e_m(\vp)$ is the sinc function which has negative
regions \cite{ka2}. 

Contrary to the Wigner functions themselves the mar\-gi\-nal distributions
\begin{equation}
  \label{eq:19}
  \int_{-\infty}^{\infty}dp W_{\phi}(q,p) =|\phi(q)|^2,~ \int_{-\infty}^{\infty}dq W_{\phi}(q,p) =|\hat{\phi}(q)|^2,
\end{equation}
where $\hat{\phi}(p)$ is the Fourier transform of $\phi(q)$,
\begin{equation}
  \label{eq:135}
  \hat{\phi}(p) = \frac{1}{2\pi \hbar}\int_{\mathbb{R}}dq\,\phi(q)\,e^{-iqp/\hbar},
\end{equation}
 are the usual non-negative  quantum mechanical ones. (Definition \eqref{eq:135} follows the convention of Ref.\ \cite{wig1}).
This is considered to be an essential requirement for the properties of Wigner functions \cite{wig1}. 

The situation is slightly more complicated for the Wigner function $V_{\psi}(\theta,p)$. Here we have
\begin{align} 
\int_{-\infty}^{\infty}d\bar{p}\,V_{\psi}(\theta,p)& =
  \frac{1}{2\pi}|\psi(\theta)|^2,  \label{eq:90} \\ \int_{-\pi}^{\pi}d\theta \,V_{\psi}(\theta,p)
  = \sum_{n \in \mathbb{Z}}|c_n|^2&\sinc\pi(p/\hbar-n) \equiv  \omega_{\psi}(p)\,,\label{eq:62}  \\
 \int_{-\infty}^{\infty}d\bar{p}\, \omega_{\psi}(p)\,\sinc\pi(p/\hbar&-m) = |c_m|^2,~m \in \mathbb{Z}.  
\end{align}
The last equation, which yields the quantum mechanical marginal probabilities $|c_m|^2$ for the quantized angular momentum,
follows from the previous one as a consequence of the orthonormality relations
\begin{equation}
  \label{eq:20}  
\int_{-\infty}^{\infty}d\bar{p} \, \sinc\pi(m-\bar{p})\,\sinc\pi(n-\bar{p}) = \delta_{mn}.
\end{equation} More details, especially about the interpolating role of the sinc functions, can be found in Ref.\ \cite{ka2}. \\
If the Wigner functions $W_{\phi_1}(q,p)$ and $W_{\phi_2}(q,p)$ for two states $\phi_1$ and $\phi_2$ are known, then one can
calculate the probability for the transitions $\phi_1 \leftrightarrow \phi_2$:
\begin{equation}
  \label{eq:24}
  |(\phi_2,\phi_1)_{\mathbb{R}}|^2 =2\pi\hbar \int_{\mathbb{R}^2}dq\,dp\, W_{\phi_2}(q,p)\,W_{\phi_1}(q,p),
\end{equation}
which implies
\begin{equation}
  \label{eq:25}
   \int_{\mathbb{R}^2}dq\,dp\,W_{\phi}^2(q,p) = \frac{1}{2\pi\hbar}.
\end{equation}
An analogue relation hold for the cylindrical case \cite{ka2}:
\begin{equation}\label{eq:26}
|(\psi_2,\psi_1)_{\mathbb{S}^1}|^2 = 2\pi \int_{-\infty}^{\infty}d\bar{p}\int_{-\pi}^{\pi}d\theta\,V_{\psi_2}(\theta,p)\,V_{\psi_1}(\theta,p),
\end{equation}
and therefore
\begin{equation}
  \label{eq:27}
 \int_{-\infty}^{\infty}d\bar{p}\int_{-\pi}^{\pi}d\theta\,V_{\psi}^2 (\theta,p) = \frac{1}{2\pi}.
\end{equation}
\subsubsection{Traces of operators}
It is possible to discuss the trace of the product of two Hilbert space operators in terms of the functions \eqref{eq:9} or \eqref{eq:14}
without having treated the operators themselves:

 If $A$ and $B$ are two operators in a Hilbert space with the scalar product \eqref{eq:7},
then
\begin{equation}
  \label{eq:28}
  \tr(A\cdot B) = \frac{1}{2\pi \hbar} \int_{\mathbb{R}^2}dq\,dp\,\tilde{A}(q,p)\,\tilde{B}(q,p).
\end{equation}
This relation can be obtained by using the expressions \eqref{eq:9} or \eqref{eq:11} under the last integral (see, e.g., Ref.\ \cite{groo}).

If $A$ is a density operator $\rho$ and $B$ a self-adjoint observable $O$ we get from the last equation
the expectation value of $O$ with respect to $\rho$:
\begin{equation}
  \label{eq:29}
  \langle O\rangle_{\rho} = \tr(\rho\cdot O) = \frac{1}{2\pi \hbar} \int_{\mathbb{R}^2}dq\,dp\,\tilde{\rho}(q,p)\,\tilde{O}(q,p),
\end{equation}
with, according to Eq.\ \eqref{eq:9},
\begin{equation}
  \label{eq:30}
 \tilde{ \rho}(q,p) = \int_{-\infty}^{\infty}d\xi e^{-ip\xi/\hbar}\langle q+\xi/2|\rho|q-\xi/2\rangle.\end{equation}
Inserting for $\rho$ the projection operator $P_{\phi}  = |\phi\rangle \langle \phi|$ 
the right hand side becomes $2\pi\hbar$ times the Wigner function $W_{\phi}(q,p)$ from Eq.\ \eqref{eq:6}.  

Thus, 
\begin{equation}
  \label{eq:31}
  W_{\rho}(q,p) = \frac{1}{2\pi \hbar}\tilde{ \rho}(q,p) = \frac{1}{2\pi \hbar}  \tr[\rho\, \Delta(q,p)]\end{equation}
is the generalization of the Wigner function from that for a pure state to that of a mixed state!

Another useful relation is obtained for $A = \rho_1$ and $B = \rho_2$:
\begin{equation}
  \label{eq:32}
  \tr(\rho_1\cdot \rho_2) = 2\pi \hbar \int_{\mathbb{R}^2}dq\,dp\,W_{\rho_1}(q,p)\,W_{\rho_2}(q,p).
\end{equation}
If $\rho_2 = \rho_1 = \rho$ then
\begin{equation}
  \label{eq:33}
  2\pi \hbar \int_{\mathbb{R}^2}dq\,dp\,W_{\rho}^2(q,p) =\tr(\rho^2) \leq 1.
\end{equation}

Relations completely analogue to those in Eqs.\ \eqref{eq:28}-\eqref{eq:33} above  are derived in Ref.\ \cite{ka2}
for the cylindrical case (note the difference in normalization: $\tr[\Delta(q,p)] = 1$, $\tr[V(\theta,p)] =1/(2\pi)$):
\begin{equation}
  \tr(A\cdot B) = 2\pi \int_{-\infty}^{\infty}d\bar{p}\int_{-\pi}^{\pi}d\theta\,\label{eq:34}
\tr[A\cdot V(\theta,p)]\,
\tr[B\cdot V(\theta,p)]\,.
\end{equation}
Applied to the expectation value of an operator $O$
for a given density operator $\rho$ gives
\begin{eqnarray} \langle O\rangle_{\rho} &=& \tr(\rho\cdot O)\label{eq:35} \\
 & =& 2\pi \int_{-\infty}^{\infty}d\bar{p}\int_{-\pi}^{\pi}d\theta\,
\tr[\rho\cdot V(\theta,p)] \nonumber
\tr[O\cdot V(\theta,p)]\,,
\end{eqnarray}
Here $V_{\rho}(\theta,p) =\tr[\rho\cdot V(\theta,p)]$ is the Wigner function for a given $\rho$.

In addition we have
\begin{equation}
  \label{eq:36}
  \int_{-\infty}^{\infty}d\bar{p}\int_{-\pi}^{\pi}d\theta
\,V_{\rho_2}(\theta,p)\,
V_{\rho_1}(\theta,p) = \frac{1}{2\pi}\tr(\rho_2\cdot \rho_1)\,.
\end{equation}
Other properties of $V_{\rho}(\theta,p)$ are discussed in Ref.\ \cite{ka2}
\subsubsection{Recovering the wave function}
If the Wigner functions $W_{\phi}(q,p)$ or $V_{\psi}(\theta,p)$ of wave functions $\phi(x)$ or $\psi(\vp)$
are given, then the wave functions  may be retrieved up to an over all constant phase \cite{coh}:
Multiplying the Wigner function \eqref{eq:6} by $\exp(ip\xi_1/\hbar)$ and integrating over $p/\hbar$ 
yields a delta funcction $\delta(\xi-\xi_1)$ which leads to
\begin{equation}
  \label{eq:37}
  \phi^{\ast}(q-\xi_1/2)\phi(q+\xi_1/2) = \int_{-\infty}^{\infty}d(p/\hbar)\,e^{ip\xi_1/\hbar}W_{\phi}(q,p).
\end{equation}
Finally, putting first $q=\xi_1/2$ and afterwards renaming $\xi_1$ by  $q$ gives
\begin{equation}
  \label{eq:38}
  \phi^{\ast}(0)\,\phi(q) = \int_{-\infty}^{\infty}d(p/\hbar)\,e^{ipq/\hbar}W_{\phi}(q/2,p).
\end{equation}
Here $\phi^{\ast}(0)$ ist a fixed complex number the modulus of which can be absorbed into the
normalization of $\phi(q)$ and the constant phase of which has no physical significance.

Exactly the same reasoning leads in the cylindrical case to the result
\begin{equation}
  \label{eq:39}
  \psi^{\ast}(0)\,\psi(\theta) = 2\pi\int_{-\infty}^{\infty}d\bar{p}\,e^{ip\theta/\hbar}V_{\psi}(\theta/2,p).
\end{equation}
\subsection{Moyal functions}
The Wigner functions \eqref{eq:6} and \eqref{eq:17} are special cases of bilinear forms for
wave functions in which both wave functions $\psi_1(\vp)$ and $\psi_2(\vp)$ or $\phi_1(x)$ and $\phi_2(x)$
respectively are identified. The general forms are
\begin{equation}
  \label{eq:41}
  W_{\phi_2\phi_1}(q,p) = \frac{1}{2\pi\hbar}\int_{-\infty}^{\infty}d\xi\,e^{-ip\xi/\hbar}\phi_2^{\ast}(q-\xi/2)\phi_1(q+\xi/2),
\end{equation}
and
\begin{equation}
  \label{eq:42}
 V_{\psi_2\psi_1}(\theta,p) =
 \frac{1}{2\pi}\int_{-\pi}^{\pi}\frac{d\vp}{2\pi}\,e^{-ip\vt/\hbar}\psi_2^{\ast}(\theta-\vt/2)\psi_1(\theta+\vt/2). 
\end{equation}
These so-called Moyal functions \cite{moy} have a number of interesting properties of which a few will be
listed here. We start with $W_{\phi_2\phi_1}(q,p)$ relying mainly on Refs.\ \cite{fol,gos}:
\begin{gather}
\int_{-\infty}^{\infty}dp\,W_{\phi_2\phi_1}(q,p) = 
\phi_2^{\ast}(q)\, \phi_1(q)\,, \label{eq:54}\\
\int_{-\infty}^{\infty}dq\,W_{\phi_2 \phi_1}(q,p) = \hat{\phi}_2^{\ast}(p)\,\hat{\phi}_1(p)\,, \label{eq:53}\\ 
\int_{-\infty}^{\infty}dp\int_{-\infty}^{\infty}dq
\,W_{\phi_2 \phi_1}(q,p) = (\phi_2,\phi_1)\,,\label{eq:80} \\
\int_{-\infty}^{\infty}dp\int_{-\infty}^{\infty}dq \, W^{\ast}_{\phi_2 \phi_1}(q,p)\,
W_{\phi_4 \phi_3}(q,p) = \label{eq:45} \\ = \frac{1}{2\pi\hbar}
(\phi_1,\phi_3)\,(\phi_2,\phi_4)^{\ast}\,.\nonumber
\end{gather}
For the cylindrical case we have \cite{ka2}
\begin{gather}
\int_{-\infty}^{\infty}d\bar{p}\,V_{\psi_2\psi_1}(\theta,p) = 
\frac{1}{2\pi}\psi_2^{\ast}(\theta)\, \psi_1(\theta)\,, \label{eq:52}\\
\int_{-\pi}^{\pi}d\theta\,V_{\psi_2 \psi_1}(\theta,p) = \sum_{m \in
    \mathbb{Z}}c_m^{(2)\ast}\sinc\pi(p/\hbar-m)c_m^{(1)}\,, \label{eq:51}\\ 
\int_{-\infty}^{\infty}d\bar{p}\int_{-\pi}^{\pi}d\theta
\,V_{\psi_2 \psi_1}(\theta,p) = (\psi_2,\psi_1)\,,\label{eq:43} \\
\int_{-\infty}^{\infty}d\bar{p}\int_{-\pi}^{\pi}d\theta \, V^{\ast}_{\psi_2 \psi_1}(\theta,p)\,
 V_{\psi_4 \psi_3}(\theta,p) = \label{eq:46} \\ = \frac{1}{2\pi}
(\psi_1,\psi_3)\,(\psi_2,\psi_4)^{\ast}\,.\nonumber
\end{gather}
Like for the Eqs.\ \eqref{eq:19} and \eqref{eq:62},  the main difference between both cases can be seen from the relations \eqref{eq:53} and \eqref{eq:51}:
the classical angular momentum $p$ is continuous whereas the quantum mechanical one is discrete and the sinc function provides the interpolation.

Again using the relation \eqref{eq:20} yields the analogue to Eq.\ \eqref{eq:53} ($n$ correponds to $p$) :
\begin{gather}
\int_{-\infty}^{\infty}d\bar{p} \, \sinc\pi(n-\bar{p}) \label{eq:55} 
 \int_{-\pi}^{\pi}d\theta\,V_{\psi_2 \psi_1}(\theta,p)  \\
= c_n^{(2)\ast}\,c_n^{(1)}\,. \nonumber
\end{gather}
\section{Operators from their\\ phase space function}
\subsection{The cylindrical phase space}
In Eqs. \eqref{eq:11} and \eqref{eq:14} we have seen how a phase space function can
be associated with a given Hilbert space operator, for the planar case and for the cylindrical
one as well. 

There is, of course, the question: given phase space functions $\tilde{A}(q,p)$ or $\tilde{A}(\theta,p)$,
how can one construct the corresponding Hilbert space operators? The answer is well-known
for a planar phase space (see below) and will now first be discussed  for the cylindrical one.
Here it is sufficient to determine all matrix elements $A_{mn} =(e_m,Ae_n)$ of an operator
$A$ (we drop the index $"\mathbb{S}^1"$ of the scalar product \eqref{eq:1} in this subsection).
Essential use will be made of the important relation \cite{ka2}
\begin{equation}
  \label{eq:58}
   \int_{-\pi}^{\pi}d\theta\int_{-\infty}^{\infty}d\bar{p}\, V_{k l}(\theta,p)\, V_{m n}(\theta,p)
 =\delta_{kn}\delta_{lm}. 
\end{equation}
In the present subsection this relation will  be used more consequently than in Subsec.\ IV B of Ref. \cite{ka2}.
\subsubsection{Single Operators}
Using the expansion \eqref{eq:3}, Eqs.\ \eqref{eq:126} and \eqref{eq:58} we get
\begin{gather}
(\psi_2,A\psi_1) =\sum_{kl}c_k^{(2)\ast} A_{kl}\,c_l^{(1)} \label{eq:59} \\
 = 2\pi\int_{-\infty}^{\infty}d\bar{p}\int_{-\pi}^{\pi}d\theta\,\sum_{klmn}
c_m^{(2)\ast} V_{mn}(\theta,p)c_n^{(1)} A_{kl}V_{lk}(\theta,p)\nonumber \\ \nonumber
=  \int_{-\infty}^{\infty}d\bar{p}\int_{-\pi}^{\pi}d\theta\, V_{\psi_2\psi_1}(\theta,p)\,\tilde{A}(\theta,p),\\
 \tilde{A}(\theta,p) = 2\pi \tr[A\cdot V(\theta,p)],~ V(\theta,p) =(V_{mn}(\theta,p)), \nonumber
\end{gather}
 implying
\begin{equation}
  \label{eq:60}
 A= (A_{mn}) = \int_{-\infty}^{\infty}d\bar{p}\int_{-\pi}^{\pi}d\theta\, V(\theta,p)\,\tilde{A}(\theta,p).
\end{equation}
This formula allows to calculate the matrix elements $A_{mn}$ once the function $\tilde{A}(\theta,p)$
is given.

A simple example: $\tilde{A}(\theta,p)= p^2$ (see Eq.\ \eqref{eq:136}). Integrating $V_{mn}(\theta,p)$
in Eq.\ \eqref{eq:4} over $\theta$ gives a factor $2\pi\delta_{mn}$. It remains
\begin{eqnarray}
  \label{eq:61}
A_{mn}&=&  \delta_{mn}\int_{-\infty}^{\infty}d\bar{p}\,p^2\int_{-\pi}^{\pi}\frac{d\vt}{2\pi}\,e^{-ip\vt/\hbar}e^{im\vt}\\\nonumber
&=&  \delta_{mn}\int_{-\infty}^{\infty}d\bar{p}\,\int_{-\pi}^{\pi}\frac{d\vt}{2\pi}\,[(-\hbar^2\partial_{\vt}^2)\,e^{-ip\vt/\hbar}]e^{im\vt}\\
&=& \delta_{mn}\int_{-\pi}^{\pi}d\vt\,[(-\hbar^2 \,\partial^2_{\vt}\delta(\vt)]e^{im\vt}\nonumber \\
&=& \hbar^2\,m^2\,\delta_{mn} =(e_m,L^2e_n),
\end{eqnarray}
with $L$ from Eq.\ \eqref{eq:15}. Here again the relation \eqref{eq:118} has been used.
\subsubsection{Product of operators I: convolution}
Of considerable interest is  the correspondence between the product $A\cdot B$ of two Hilbert space operators 
and their phase space function $\widetilde{A B}(\theta)$ (the usual ``dot'' between
the operators is omitted for their phase space functions): According to the general Eq.\ \eqref{eq:60} we have
\begin{equation}
  \label{eq:107}
  A\cdot B =\int_{-\infty}^{\infty}d\bar{p}\int_{-\pi}^{\pi}d\theta\, V(\theta,p)\,\widetilde{A B}(\theta,p).
\end{equation}
Again using the relation \eqref{eq:58} one can verify that
\begin{gather}
  \label{eq:108}
  \widetilde{A B}(\theta,p) = 2\pi \tr[V(\theta,p) \cdot A\cdot B] \\ =\int_{-\infty}^{\infty}d\bar{p}_1\int_{-\pi}^{\pi}d\theta_1\, 
G_A(\theta,p;\theta_1,p_1)\tilde{B}(\theta_1,p_1), \nonumber \\
G_A(\theta,p\,;\theta_1,p_1)= 2\pi \tr[V(\theta,p)\cdot A\cdot V(\theta_1,p_1)]. \nonumber
\end{gather}
Here the phase space function of the product $A\cdot B$ is a kind of convolution of those for the single operators. Compare Ref.\ \cite{wig3}
for a similar relation in the planar case.
\subsubsection{Product of operators II: $\star$ product}
It turns out that  the procedure which is employed in the planar case \cite{groo,wig1,fol,gos,zac,case,zac2} can be used in the cylindrical one, too:

Starting again with the relation
\begin{equation}
  \label{eq:115}
   \widetilde{A B}(\theta,p) = 2\pi \tr[V(\theta,p) \cdot A\cdot B],
\end{equation}
and using for $A$ and $B$ inside the trace the representation \eqref{eq:60} yields
\begin{gather}
  \label{eq:116}
  \widetilde{A B}(\theta,p)  = 2\pi \int_{-\infty}^{\infty}d\bar{p}_1\,d\bar{p}_2\int_{-\pi}^{\pi}d\theta_1\,d\theta_2  \\ \times
\tr[V(\theta,p)\cdot V(\theta_1,p_1)\cdot V(\theta_2,p_2)] \nonumber \\ \times
\tilde{A}(\theta_1,p_1)\,\tilde{B}(\theta_2,p_2). \nonumber
\end{gather}
Inserting for $V$ the integral representation \eqref{eq:4} and using
 relation \eqref{eq:118}
  one obtains for the trace:
\begin{gather}
  \label{eq:117}
\tr[V(\theta,p)\cdot V(\theta_1,p_1)\cdot V(\theta_2,p_2)] = \\
= \frac{4}{(2\pi)^3}e^{-2i[p(\theta_1-\theta_2) + p_1(\theta_2-\theta) + p_2(\theta-\theta_1]/\hbar}.  
\end{gather}
Observing that
\begin{align}
  \label{eq:119}
 & [p(\theta_1-\theta_2) + p_1(\theta_2-\theta) + p_2(\theta-\theta_1] = \\
&= -[(\theta_1-\theta)b - \alpha(p_1-p)], \nonumber \\
&~~~~~ \alpha = \theta_2-\theta,~~b= p_2-p,~~d\theta_2\,dp_2 = d\alpha\,db, \nonumber\\
& \tilde{B}(\theta + \alpha, p+b) = e^{(\alpha\partial_{\theta} + b\partial_p)}\tilde{B}(\theta,p), \nonumber \\
& \alpha\,e^{-2i\alpha(p_1-p)/\hbar} = -\frac{i\hbar}{2}\partial_pe^{-2i\alpha(p_1-p)/\hbar},\nonumber \\
& b\,e^{2ib(\theta_1-\theta)/\hbar} = \frac{i\hbar}{2}\partial_{\theta}e^{2ib(\theta_1-\theta)/\hbar} \nonumber
\end{align}
yields
\begin{gather}
  \label{eq:120}
  \widetilde{A B}(\theta,p)  = \frac{4}{(2\pi)^2} \int_{-\infty}^{\infty}d\bar{p}_1\,d\bar{b}\int_{-\pi}^{\pi}d\theta_1\,d\alpha  \\ \times
\tilde{A}(\theta_1,p_1) \{e^{2i[(\theta_1-\theta)b -(p_1-p)\alpha]/\hbar} e^{\frac{\hbar}{2i}\Lambda(\theta,p)}\tilde{B}(\theta,p)\},\\
\Lambda(\theta,p) = \stackrel{\leftarrow}{\partial}_p\,\stackrel{\rightarrow}{\partial}_{\theta}
- \stackrel{\leftarrow}{\partial}_{\theta}\,\stackrel{\rightarrow}{\partial}_p, 
\end{gather}
where $\stackrel{\leftarrow}{\partial}_p$ and $\stackrel{\leftarrow}{\partial}_{\theta}$ act to the left, on the exponentials.

Integration over $\bar{b}$ gives a delta function $\delta(\theta_1-\theta)$ which allows to carry out the $\theta_1$-integration.
It remains the integral
\begin{equation}
  \label{eq:121}
  \frac{1}{\pi} \int_{-\infty}^{\infty}d\bar{p}_1\int_{-\pi}^{\pi}d\alpha \tilde{A}(\theta,p_1)\,e^{-2i(p_1-p)\alpha/\hbar}.
\end{equation}
Inserting into 
\begin{equation}
  \label{eq:122}
  \tilde{A}(\theta,p_1) = 2\pi \tr(A\cdot V(\theta,p_1),
\end{equation}
for $V(\theta,p_1)$ the representation \eqref{eq:4} 
and carrying out the integrations gives for the integral \eqref{eq:121}
\begin{equation}
  \label{eq:123}
   \frac{1}{\pi} \int_{-\infty}^{\infty}d\bar{p}_1\int_{-\pi}^{\pi}d\alpha \tilde{A}(\theta,p_1)\,e^{-2i(p_1-p)\alpha/\hbar} = \tilde{A}(\theta,p),
\end{equation}
so that finally
\begin{eqnarray}
  \label{eq:124}
  \widetilde{AB}(\theta,p)& =& \tilde{A}(\theta,p)\,e^{\frac{\hbar}{2i}\Lambda(\theta,p)}\tilde{B}(\theta,p)\\
                            &\equiv & \tilde{A}(\theta,p)\star\tilde{B}(\theta,p)\\
                          &=&\tilde{B}(\theta,p)\,e^{-\frac{\hbar}{2i}\Lambda(\theta,p)}\tilde{A}(\theta,p). \nonumber
\end{eqnarray}
This is the complete analogue to the corresponding formula in the planar case (see the next subsection). Thus, the $\star$ product
formalism of the planar phase space \cite{groen,imr,bay1,bay2,fron,tak,fair2,hir,kon,han,fair3,fre} can be carried over to the cylindrical one! Only a few important example
will be mentioned here:

From Eq.\ \eqref{eq:124} one obtains for commutator and antcommutator:
\begin{eqnarray}
  \label{eq:145}
  \widetilde{[A,B]}(\theta,p)& =& 2i\tilde{A}(\theta,p)\sin[\frac{\hbar}{2}\Lambda(\theta,p)]\tilde{B}(\theta,p), \\
\widetilde{\{A,B\}}(\theta,p)& =& 2\tilde{A}(\theta,p)\cos[\frac{\hbar}{2}\Lambda(\theta,p)]\tilde{B}(\theta,p).\label{eq:146}
\end{eqnarray}

In first order of $\hbar$ the right hand side of Eq.\ \eqref{eq:145} becomes - as usual - $i\hbar$ times the Poisson bracket of the functions $\tilde{A}$
and $\tilde{B}$ and the right hand side of \eqref{eq:146} becomes in zeroth order twice their product.

The expression for the commutator is of special importance for the time evolutions of an operator $A$ or that of a Wigner function which are
determined by the commutator $[H,A]$ etc. (see Sec.\ IV below).

Eq.\ \eqref{eq:124} can also be written in terms of the Bopp operators \cite{bopp,bop} which are especially helpful for explicit calculations:
\begin{equation}
  \label{eq:144}
  \widetilde{A\cdot B}(\theta,p) =  \tilde{A}(\theta,p)\star\tilde{B}(\theta,p) =  \tilde{A}(\tilde{\theta},\tilde{p})\cdot \tilde{B}(\theta,p),
\end{equation}
where 
\begin{equation}
  \label{eq:143}
  \tilde{\theta} = \theta - \frac{\hbar}{2i}\partial_p,~~\tilde{p} = p+ \frac{\hbar}{2i}\partial_{\theta}.
\end{equation}

As a simple application of Eq.\ \eqref{eq:144} consider the example discussed in Subsec.\ II A above: $
\tilde{L}(\theta,p) = p$ and $\tilde{C}(\theta,p) = \cos\theta$: Using Eq.\ \eqref{eq:144} we get
\begin{equation}
  \label{eq:147}
  \widetilde{LC}(\theta,p) = (p + \frac{\hbar}{2i}\partial_{\theta})\cos\theta = p\cos\theta - \frac{\hbar}{2i}\sin\theta 
\end{equation}
and
\begin{gather}
  \label{eq:148}
\widetilde{CL}(\theta,p) =
  \cos(\theta - \frac{\hbar}{2i}\partial_p)\,p=\\ = [\cos \theta\,\cos(\frac{\hbar}{2i}\partial_p) + \sin\theta\,\sin(\frac{\hbar}{2i}\partial_p)]\,p \nonumber \\
= p\cos\theta + \frac{\hbar}{2i}\sin\theta, \nonumber
\end{gather}
which coincides with Eqs.\ \eqref{eq:128} and \eqref{eq:149}.

In both cases, \eqref{eq:147} and \eqref{eq:148}, the relation
\begin{equation}
  \label{eq:156}
  \cos(x-y) = \cos x\,\cos y + \sin x\,\sin y
\end{equation}
has been used.
\subsection{The planar phase space}
We recall only  very briefly a few essential elements of constructing Hilbert space operators from functions on the planar phase space,
namely mainly those, which correspond closely to relations of the cylindrical case discussed in the last subsection. Note 
the many references  relating to the planar case given there.

Like the operator $V(\theta,p)$ in the cylindrical case the operator $\Delta(q,p)$ from Eq.\ \eqref{eq:10} can play a crucial role for the planar one:
Its importance for the Wigner function on the planar phase space was emphasized by Leaf \cite{leaf1,leaf2}. De Groot and Suttorp
gave a  very recommendable  exposition of that approach in their textbook \cite{groo}. 

The inversion of  relation \eqref{eq:11} is
\begin{equation}
  \label{eq:66}
  A = \frac{1}{2\pi\hbar}\int_{-\infty}^{\infty}\int_{-\infty}^{\infty}dq\,dp\,\Delta(q,p)\,\tilde{A}(q,p),
\end{equation}
which is the analogue to Eq.\ \eqref{eq:60}.

As $\Delta(q,p)$ can also be written as
\begin{equation}
  \label{eq:67}
 \Delta(q,p)= \frac{1}{2\pi\hbar}\int_{-\infty}^{\infty}\int_{-\infty}^{\infty}du\,dv\,e^{-i[(Q-q)u +(P-p)v]/\hbar},
\end{equation}
where $Q$ and $P$ are the usual position and momentum operators, 
the operator $A$ may also be (Weyl) represented as
\begin{equation}
  \label{eq:68}
  A = \int_{-\infty}^{\infty}\int_{-\infty}^{\infty}du\,dv\,e^{-i(Q u + Pv]/\hbar} \hat{A}(u,v),
\end{equation}
where $\hat{A}(u,v)$ is the Fourier transform of $\tilde{A}(q,p)$:
\begin{equation}
  \label{eq:69}
  \tilde{A}(q,p) = \int_{-\infty}^{\infty}\int_{-\infty}^{\infty}du\,dv\,\hat{A}(u,v) e^{-i(q u + pv)/\hbar}.
\end{equation}
The important operator $\Delta(q,p)$ is hardly mentioned or used in most of the literature on the
planar Wigner function. Probably the main reason is that the above formulas can be rewritten by using the
operator relations
\begin{equation}
  \label{eq:70}
  [P,Q]= \frac{\hbar}{i},~~ e^{A+B}=e^A\cdot e^B\cdot e^{-[A,B]/2}.
\end{equation}
They lead to
\begin{eqnarray}
  \label{eq:71}
  A& =& \frac{1}{(2\pi\hbar)^2}\int_{-\infty}^{\infty}dq\,dp\,du\,dv\,\tilde{A}(q,p)\times \\ && \times e^{iuv/(2\hbar)} e^{-i(Q-q)u/\hbar}\cdot e^{-i(P-p)v/\hbar}.
\nonumber \end{eqnarray}
Replacing the product $uv$ by the differential operators $-\hbar^2\partial_q\,\partial_p$ and integrating partially one obtains
\begin{eqnarray}
  \label{eq:72}
   A& =& \frac{1}{(2\pi\hbar)^2}\int_{-\infty}^{\infty}dq\,dp\,du\,dv \times \\ && \times
 e^{-i(Q-q)u/\hbar}\cdot e^{-i(P-p)v/\hbar}\,\exp[\frac{\hbar}{2i}\partial_q\partial_p]\,\tilde{A}(q,p). \nonumber
\end{eqnarray}
Here the operator ordering is such that the position operators are to the left of the momentum ones.

The phase space function $\widetilde{AB}(q,p)$ of the product $A\cdot B$ of two operators is
given by 
\begin{eqnarray}
  \label{eq:73}
  \widetilde{AB}(q,p)&=& \tilde{A}(q,p)\,\exp[\hbar\Lambda(q,p)/(2i)]\tilde{B}(q,p),\\
&&\Lambda(q,p) = \stackrel{\leftarrow}{\partial}_p\,\stackrel{\rightarrow}{\partial}_q
- \stackrel{\leftarrow}{\partial}_q\,\stackrel{\rightarrow}{\partial}_p. \nonumber
\end{eqnarray}
The  expression \eqref{eq:124} for the cylindrical case is the analogue of  relation \eqref{eq:73}, which was first derived by Groenewold \cite{groen}.

There are, of course, many more important properties of Wigner functions for the planar phase space,
which have been worked out over decades (see the numerous references in the last Subsec.\ IIIA). For most of them their analogues in the cylindrical case
 wait for new applications.
\section{Dynamical and energy equations}
Quantum mechanics provides - among others - two basic elements for the description of atomic systems:

First, it gives a description of their time evolution, i.e. their dynamics, either - in the Schr\"odinger picture - in terms of
a  Schr\"odinger equation  for the wave function of the system, or - in the Heisenberg picture - in terms of  equations of motion for
operators (``observables'') which characterize the system.

Second, it allows - in principle - for the calculation of eigenvalues and eigenstates of observables, especially the energy,
which are important for a quantitative description of the system. In addition, in the realm of continuous (higher) energies
it provides the basis for the description of, e.g., stationary scattering processes for a given energy.

In this second case, which concerns the stationary structure of physical systems, the time-independent Schr\"odinger equation
is an important tool.

Similarly one can ask, how  the time evolution of a system, from a given initial state, is described by its associated Wigner function and:  are there
appropriate structure equations corresponding to the eigenvalue etc. equations of the conventional time-independent quantum mechanics.

The question of time evolution was already answered by Wigner in his very first paper on the subject for planar phase spaces \cite{wig}
in terms of a generalized Liouville equation. The problem was later discussed extensively by Moyal \cite{moy}. For the further development of this subject
see the reviews \cite{taka,wig1,lee,schl,dra}.

 The discussion of time-independent structure or energy ``eigenvalue'' equations for Wigner functions has its origin in applications to statistical
 mechanics systems, the topic of
Wigner's seminal paper \cite{wig}: For a quantum canonical ensemble in equilibrium the quantum mechanical time evolution operator
\begin{equation}
  \label{eq:138}
  U(t) = e^{-itH/\hbar}
\end{equation}
is  replaced by 
\begin{equation}
  \label{eq:139}
 \Omega(\beta) = e^{-\beta H},~\beta = \frac{1}{k_B T},
\end{equation}
where $H$ is the associated Hamilton operator.
The operator $\Omega$ provides the density operator
\begin{equation}
  \label{eq:140}
  \rho(\beta) =\Omega(\beta)/Z(\beta),~~Z(\beta) = \tr(\Omega)
\end{equation}
of the system  and obeys the Bloch equation \cite{blo}
\begin{equation}
  \label{eq:141}
  \partial_{\beta}\Omega = -H\cdot \Omega = -\Omega \cdot H.
\end{equation}
Wigner functions related to $\Omega$ have been discussed in Refs. \cite{opp,imr}. A brief summary is contained in Ref. \cite{wig1}.

Comparing Eqs. \eqref{eq:138} and \eqref{eq:139} we note the correspondence
\begin{equation}
  \label{eq:142}
  -t/\hbar \longleftrightarrow i\beta.
\end{equation}
In modern approaches \cite{haa,bor,thi} this relation is treated under the topic ``KMS condition'' which deals with the analytic continuation between the
two regions in the complex $t$-plane.
\subsection{Time evolution of the Wigner function \\ for a density matrix}
For the time dependence of a density operator $\rho(t)$ von Neumann's equation holds:
\begin{equation}
  \label{eq:150}
  i\hbar\partial_t \rho(t) = [H,\rho(t)].
\end{equation}
In order to translate this into the corresponding equation for the phase space function 
 \begin{equation}
   \label{eq:152}
   \tilde{\rho}(\theta,p;t) = 2\pi V_{\rho}(\theta,p;t) =
 2\pi \tr[\rho(t)\cdot V(\theta,p)]
 \end{equation}
we need the phase space function $\tilde{H}(\theta,p)$ of the Hamilton operator \eqref{eq:21}. According to Eq.\ \eqref{eq:151}
this is given by
\begin{equation}
  \label{eq:153}
  \tilde{H}(\theta,p) = \gamma\,p^2 + U(\theta).
\end{equation}
In the following we use for $U(\theta)$ the pendulum potential
\begin{equation}
  \label{eq:154}
  U(\theta) = -A\,\cos\theta
\end{equation} 
and for $\rho(t)$ the diagonal form \cite{ka2}
\begin{align}
  \label{eq:155}
  \rho(t) =& (\rho_{mn}(t)),~~\rho_{mn}(t) =\lambda_m(t)\,\delta_{mn},\\
&\lambda \geq 0,~~\sum_{m \in \mathbb{Z}}\lambda_m(t) =1. \nonumber
\end{align}
This gives
\begin{eqnarray}
  \label{eq:157}
  V_{\rho}(\theta,p;t)& =& \frac{1}{2\pi}\sum_{m \in \mathbb{Z}}\lambda_m(t)\sinc\pi(p/\hbar -m)~~~~,\\
\sinc\pi(p/\hbar -m)& =& \frac{1}{2\pi}\int_{-\pi}^{\pi}d\vt e^{-i(p/\hbar -m)\vt}. \label{eq:158}
\end{eqnarray}
Thus, $V_{\rho}(\theta,p;t)$ is independent of $\theta$.

The more general $\theta$-dependent case will be treated in the next subsection, where additional remarks on the pendulum can be found, too.

Using relations \eqref{eq:144}, \eqref{eq:143} and \eqref{eq:156} we have
\begin{align}
  \label{eq:159}
 & (p + \frac{\hbar}{2i}\partial_{\theta})^2\cdot  V_{\rho}(p;t)\nonumber   = p^2\, V_{\rho}(p;t), \nonumber \\ &
\cos(\theta - \frac{\hbar}{2i}\partial_{p}) V_{\rho}(p;t) = [\cos\theta\,\cos(\frac{\hbar}{2i}\partial_{p})\nonumber \\ &  +
 \sin\theta \sin(\frac{\hbar}{2i}\partial_{p})]V_{\rho}(p;t),\nonumber  \\
\widetilde{HV_{\rho}}(\theta,p)& = \{\gamma\,p^2 -A[\cos\theta\,\cos(\frac{\hbar}{2i}\partial_{p})\\ &  + \sin\theta  \sin(\frac{\hbar}{2i}\partial_{p})]\}V_{\rho}(p;t),
 \nonumber
\end{align}
and
\begin{align}
  \label{eq:160}
 & \sinc\pi[p/\hbar + \frac{1}{2i}\partial_{\theta}-m]\cos\theta \nonumber \\ & = \frac{1}{2\pi}\int_{-\pi}^{\pi}d\vt e^{-i[p/\hbar -m +\frac{1}{2i}\partial_{\theta}]\vt}\cos\theta \nonumber \\ &
= \frac{1}{2\pi}\int_{-\pi}^{\pi}d\vt\, \cos(\theta -\vt/2) e^{-i(p/\hbar -m)\vt}\nonumber \\ & =[\cos\theta\,\cos(\frac{\hbar}{2i}\partial_{p})-\nonumber \\ &-\sin\theta
 \sin(\frac{\hbar}{2i}\partial_{p})]\sinc\pi(p/\hbar -m),\nonumber \\ 
\widetilde{V_{\rho}H}(\theta, p)& =\{\gamma\,p^2 -A[\cos\theta\,\cos(\frac{\hbar}{2i}\partial_{p}) \\ &
 - \sin\theta \sin(\frac{\hbar}{2i}\partial_{p})]\}V_{\rho}(p;t), \nonumber
\end{align}
so that
\begin{equation}
  \label{eq:161}
  \widetilde{[H,V_{\rho}]} = -2A \sin\theta\,[ \sin(\frac{\hbar}{2i}\partial_{p})]V_{\rho}(p;t).
\end{equation}
Combined with Eq.\ \eqref{eq:150} this yields
\begin{equation}
  \label{eq:162}
  \partial_tV_{\rho}(p;t) = -\frac{2A}{i\hbar}\sin\theta\,\sin(\frac{\hbar}{2i}\partial_{p})\,V_{\rho}(p;t).
\end{equation}
With
\begin{equation}
  \label{eq:92}
  \sin z = \sum_{n=0}^{\infty}(-1)^n \frac{z^{2n+1}}{(2n+1)!}
\end{equation}
and
\begin{equation}
  \label{eq:163}
  A\,\sin\theta = \partial_{\theta}U(\theta)
\end{equation}
 Eq.\ \eqref{eq:162} can be written as
\begin{align}
  \label{eq:164}
 & [\partial_t -\partial_{\theta}U(\theta)\,\partial_p]V_{\rho}(p;t)\\& =
 \partial_{\theta}U(\theta)\,[\sum_{n=1}^{\infty}(\frac{\hbar}{2})^{2n}\,
\frac{\partial_p^{2n+1}}{(2n+1)!}]V_{\rho}(p;t). \nonumber
\end{align}
Note that the left hand side of the last equation is independent of $\hbar$, whereas the right hand side represents quantum effects.

\subsection{Generalized Liouville equation} 
The time evolution of the model discussed in the last subsection is an example (see the left hand side of Eq.\ \eqref{eq:161}) for 
the general time evolution equation, derived in Ref.\ \cite{ka2}, for a Moyal function $V_{\psi_2\psi_1}(\theta,p;t)$:
\begin{eqnarray}
  \label{eq:74}
\partial_tV_{\psi_2\psi_1}(\theta,p;t)& =&\frac{1}{\hbar}(\psi_2(t),K(\theta,p)\psi_1(t))_{\mathbb{S}^1}, \\
K(\theta,p)& =& i[H,V(\theta,p)].
\end{eqnarray}
In the following the right Hand side of this equation  will be worked out in detail directly, partially following Ref.\ \cite{schl}.

 We start from
\begin{eqnarray}
  \label{eq:75}
  i\hbar \partial_t\psi(\theta;t)& =& H\psi(\theta;t),~H=H_0+U(\theta),\\\label{eq:114}
H_0& =& \frac{1}{2mr_0^2}\,L^2 =-\varepsilon\frac{d^2}{d\theta^2},~\varepsilon= \frac{\hbar^2}{2mr_0^2},~~~~~
\end{eqnarray} where
$mr_0^2$ is the moment of inertia of a point mass rotating at a distance $r_0$ around an axis. \\
In the following we abbreviate:
\begin{equation}
  \label{eq:76}
  \psi_2(\theta -\vt/2) \equiv \psi_2,~~\psi_1(\theta + \vt/2) \equiv \psi_1.
\end{equation}
It follows from Eqs.\ \eqref{eq:42} and \eqref{eq:75} that
\begin{equation}
  \label{eq:77}
   i\hbar \partial_tV_{\psi_2\psi_1} = \frac{1}{2\pi}\int_{-\pi}^{\pi} \frac{d\vt}{2\pi}e^{-ip\vt/\hbar}[-(H\psi_2^{\ast})\psi_1 +\psi_2^{\ast}(H\psi_1)].
\end{equation}
Starting with the $H_0$ part, observing that
\begin{equation}
  \label{eq:78}
  \frac{d\psi_2}{d\theta} = -2\frac{d\psi_2}{d\vt},~\frac{d\psi_1}{d\theta} = 2\frac{d\psi_1}{d\vt},
\end{equation}
and then integrating partially with respect to $\vt$ under the integral \eqref{eq:77} yields
\begin{gather}
  \label{eq:79}
 i\hbar \partial_tV_{\psi_2\psi_1}(\theta,p;t) = -\frac{2\,i\varepsilon\, p}{\hbar}\partial_{\theta}V_{\psi_2\psi_1}(\theta,p;t)- \\ \nonumber
 - \frac{2\varepsilon}{(2\pi)^2}\partial_{\theta}b_{\psi_2\psi_1}(\theta,p,\vt = \pm \pi;t),
\end{gather}
where the  boundary ``potential''
\begin{gather}
  \label{eq:109}
b_{\psi_2\psi_1}(\theta,p,\vt = \pm \pi;t)\\ =\frac{1}{(2\pi)^2}[e^{-ip\vt/\hbar}\psi_2^{\ast}(\theta-\vt/2;t)\,\psi_1(\theta+\vt/2;t)]^{\vt =\pi}_{\vt =-\pi}.
\nonumber
\end{gather}
is the difference of the values of the integrand of $V_{\psi_2\psi_1}(\theta,p)$
at the boundaries $\vt= \pm  \pi$. This term vanishes in the planar case because the integrand vanishes at infinity. It does not
in general vanish here.

 For $\psi_2(\vp) = \psi_1(\vp) = \psi(\vp)$
one can write
\begin{gather}
  \label{eq:110}
  b_{\psi}(\theta,p,\vt = \pm \pi;t)\\ =\frac{2i}{(2\pi)^2}\Im[e^{-ip\pi/\hbar}\psi^{\ast}(\theta-\pi/2)\psi(\theta+\pi/2)]. \nonumber
\end{gather}
 A simple example is
\begin{equation}
  \label{eq:81}
 \psi(\vp)  = \frac{1}{\sqrt{2}}(e^{im\vp}+ e^{in\vp}),
\end{equation}
for which the boundary term turns out to be
\begin{gather}
  \label{eq:82}
 b_{\psi}(\theta,p,\vt = \pm \pi;t)  = \frac{2i}{(2\pi)^2}\sin\pi[(m+n)/2 -p]\times \\ \times \{\cos(m-n)\theta + \cos\pi[(m-n)/2]\} \nonumber
\end{gather}
Differentiating with respect to $\theta$ (see Eq.\ \eqref{eq:79}) leaves a term proportional to
$\sin(m-n)\theta$ which vanishes for $m=n$, but not in general. It vanishes also
if $p-(m+n)/2 = k\pi,\,k \in \mathbb{Z}$. 

We next turn to the contribution of the potential $U(\theta)$ to the right hand side of Eq.\ \eqref{eq:77} (recall Eq.\ \eqref{eq:76}):
\begin{eqnarray}&& \frac{1}{2\pi}\int_{-\pi}^{\pi}\frac{d\vt}{2\pi}e^{-ip\vt/\hbar}[-U(\theta-\vt/2)\psi_2^{\ast}]\psi_1 + \psi_2^{\ast}[ U(\theta + \vt/2) \psi_1]. \nonumber \\
&& =  \frac{1}{2\pi}\int_{-\pi}^{\pi}\frac{d\vt}{2\pi}e^{-ip\vt/\hbar}[U(\theta+\vt/2) - U(\theta-\vt/2)]\psi_2^{\ast}\psi_1\label{eq:85}
\end{eqnarray}
A real periodic potential, $U(\theta + 2\pi) = U(\theta)$, can be expanded in a Fourier series
\begin{equation}
  \label{eq:83}
  U(\theta) = \sum_{k=0}^{\infty} U_k(\theta),~ U_k(\theta)= A_k\cos k\theta + B_k \sin k\theta.
\end{equation}
It follows that
\begin{gather}
  \label{eq:84}
  U_k(\theta+\vt/2)-U_k(\theta-\vt/2)\\ = 2\sin(k\vt/2) (-A_k\sin k\theta + B_k \cos k\theta)\nonumber \\
 =2 \sin(k\vt/2)\,\partial_{\theta}U_k(\theta)/k. \nonumber
\end{gather}
Inserting this into Eq.\ \eqref{eq:85} gives
\begin{equation}
  \label{eq:86}
 2 [\partial_kU(\theta)/k]\, \frac{1}{2\pi}\int_{-\pi}^{\pi}\frac{d\vt}{2\pi}e^{-ip\vt/\hbar}\sin(k\vt/2)\psi^{\ast}_2\,\psi_1.
\end{equation}
As
\begin{equation}
  \label{eq:87}
(k\vt/2)\,e^{-ip\vt/\hbar}= (k/2)i\hbar\partial_pe^{-ip\vt/\hbar},
\end{equation}
the contribution \eqref{eq:86} can be written as
\begin{equation}
  \label{eq:88}
 2 [\partial_kU(\theta)/k] \sin[(k/2)i\hbar\partial_p]V_{\psi_2\psi_1}(\theta,p),
\end{equation}
with $\sin z$ as in Eq.\ \eqref{eq:92}.

Adding the term \eqref{eq:88} to Eq.\ \eqref{eq:79} and dividing by $i\hbar$ yields
\begin{gather}
  \label{eq:89}
   \partial_tV_{\psi_2\psi_1}(\theta,p;t) = -\frac{2\varepsilon\, p}{\hbar^2}\partial_{\theta}V_{\psi_2\psi_1}(\theta,p;t)+ \\ \nonumber
+ \frac{ 2}{ik\hbar} [\partial_kU(\theta)] \sin[(k/2)i\hbar\partial_p]V_{\psi_2\psi_1}(\theta,p) \\ \nonumber
- \frac{2\varepsilon}{\hbar i}\partial_{\theta}b_{\psi_2\psi_1}(\theta,p,\vt = \pm \pi;t)
\end{gather}
Inserting $\varepsilon$ from Eq.\ \eqref{eq:114} and separating all the terms which are independent of $\hbar$ finally gives
\begin{gather}
  \label{eq:91}
  \{\partial_t + \frac{p}{mr_0^2}\partial_{\theta}-\partial_{\theta}U_k(\theta)\,\partial_p\}V_{\psi_2\psi_1}(\theta,p;t) = \\ \nonumber
= \{\partial_{\theta}U_k(\theta) \sum_{n=1}^{\infty}\frac{(k\hbar/2)^{2n} }{k(2n+1)!}\partial_p^{2n+1}\} V_{\psi_2\psi_1}(\theta,p;t) + \\ \nonumber
+ \frac{\hbar}{mr_0^2i}\partial_{\theta}b_{\psi_2\psi_1}(\theta,p,\vt = \pm \pi;t).
\end{gather}
Some remarks on the result: \\
The  generalized Liouville equation \eqref{eq:91} holds, of course, too, if $\psi_2(\vp) = \psi_1(\vp)$, i.e. for the Wigner function proper
$V_{\psi}(\theta,p;t)$:
\begin{gather}
  \label{eq:165}
   \{\partial_t + \frac{p}{mr_0^2}\partial_{\theta}-\partial_{\theta}U_k(\theta)\,\partial_p\}V_{\psi}(\theta,p;t) = \\ \nonumber
= \{\partial_{\theta}U_k(\theta) \sum_{n=1}^{\infty}\frac{(k\hbar/2)^{2n} }{k(2n+1)!}\partial_p^{2n+1}\} V_{\psi}(\theta,p;t) + \\ \nonumber
+ \frac{\hbar}{mr_0^2i}\partial_{\theta}b_{\psi}(\theta,p,\vt = \pm \pi;t),
\end{gather}
where $b_{\psi}$ is given by Eq.\ \eqref{eq:110}.

If $V_{\psi}(\theta,p;t)$ is independent of $\theta$, Eq.\ \eqref{eq:165} reduces to Eq.\ \eqref{eq:164}.

It is remarkable that   no  higher derivatives of the potential $U_k(\theta)$ enter the right hand side of Eq.\ \eqref{eq:91},
 contrary to the planar case, where the corresponding
potential term has the form \cite{schl2}
\begin{equation}
  \label{eq:137}
  \sum_{n=1}^{\infty}\frac{(-1)^n(\hbar/2)^{2n}}{(2l+1)!}[\partial_q^{2n+1}U(q)]\partial_p^{2n+1}W_{\phi}(q,p;t).
\end{equation}

The boundary ``potential'' $b_{\psi}$ on the right hand side of Eq.\ \eqref{eq:91} is proportional to $\hbar$, i.e.\ it is a quantum effect.
 In addition it makes the otherwise
homogeneous PDE an inhomogeneous one.

If the time dependence of the wave functions $\psi(\theta;t)$ in Eq.\ \eqref{eq:75} can be separated as
\begin{equation}
  \label{eq:93}
  \psi(\theta;t) = e^{-iEt/\hbar}u(\theta),
\end{equation}
then the time derivative in Eq.\ \eqref{eq:91} takes the form
\begin{equation}
\label{eq:94}
  \partial_tV_{\psi_2\psi_1}(\theta,p;t) = \frac{E_1-E_2}{i \hbar}V_{\psi_2\psi_1}(\theta,p;t).
\end{equation}
Thus,  for $\psi_2(\vp) = \psi_1(\vp)$ and a stationary Wigner function $V_{\psi}(\theta,p)$, the generalized Liouville equation \eqref{eq:165}
reduces to
\begin{gather}
  \label{eq:104}
 \{ \frac{p}{mr_0^2}\partial_{\theta}-\partial_{\theta}U_k(\theta)\,\partial_p\}V_{\psi}(\theta,p) = \\ \nonumber
= \{\partial_{\theta}U_k(\theta) \sum_{n=1}^{\infty}\frac{(k\hbar/2)^{2n} }{k(2n+1)!}\partial_p^{2n+1}\} V_{\psi}(\theta,p) + \\ \nonumber
+ \frac{\hbar}{mr_0^2i}\partial_{\theta}b_{\psi}(\theta,p,\vt = \pm \pi). 
\end{gather}
As already discussed in Subsec. IV A, an important example for a potential $U(\theta)$ of Eq.\ \eqref{eq:83} is that of a pendulum \cite{don,gre}:
\begin{equation}
  \label{eq:107}
U(\theta) = U_1(\theta) = -A\cos\theta,
\end{equation}
where $A= mg\,r_0$ for a rotating or oscillating point mass in a constant gravitational field $\vec{g}$, $A=q\,Er_0$ for such a point mass with charge $q$ in a homogeneous
electric field $\vec{E}$ and $A = p\,E$ for a (rotating) dipole $\vec{p}$ in an external electrical field $\vec{E}$.
\subsection{An energy equation for the Wigner-Moyal function}
We have seen in the last subsection that a time dependence \eqref{eq:93} for the wave functions $\psi_{1,2}(\theta;t)$ in the principal dynamical equation \eqref{eq:91}
leads to a stationary equation \eqref{eq:104} in the limit $\psi_2 \to \psi_1 = \psi$  which does not contain an energy parameter $E$ anymore. 
One can change that by two related arguments:

If one replaces $\psi_2^{\ast}(\theta;t)$ by $\psi_2^{\ast}(\theta;-t)$ then $E_2$ in Eq.\ \eqref{eq:94} changes sign. However, this operation is not the established
Wigner time inversion which involves a simultaneous complex conjugation of the wave function. Therefore, $\psi_2^{\ast}(\theta;-t)$ is equivalent to $\psi_2(\theta;t)$.
Thus, if one replaces $\psi_2^{\ast}(\theta;t)$ merely by its complex conjugate, then $E_2$ in Eq.\ \eqref{eq:94} changes sign, too. So, either of these procedures
 can make the
difference $E_1 -E_2$ in Eq.\ \eqref{eq:94} into the sum $E_1 + E_2$.

Another related argument makes use of the exchange \eqref{eq:142}: if one replaces $it/\hbar$ in the Schr\"odinger equation by $\beta$ then one achieves the same goal,
namely an equation for $E_1 + E_2$.

The discussion to obtain such an energy type equation for Wigner functions was -- implicitly --  started by Baker \cite{bak}, explicitly discussed by
Fairlie \cite{fair} and then persued by others \cite{kun,fair1,fair2,schl}.

 We will now derive the corresponding equation for the cylindrical phase space:

If we have the stationary Schr\"odinger equations
\begin{equation} 
  \label{eq:95} 
H\psi_j = E_j\psi_j,~j=1,2\,,
\end{equation}
with $H$ as in Eqs.\ \eqref{eq:75} and \eqref{eq:90}, then
\begin{gather}
  \label{eq:96}
  (E_1+E_2)V_{\psi_2\psi_1}(\theta,p)= \\ = \frac{1}{2\pi}\int_{-\pi}^{\pi} \frac{d\vt}{2\pi}e^{-ip\vt/\hbar}[(H\psi_2^{\ast})\psi_1 +\psi_2^{\ast}(H\psi_1)].\nonumber
\end{gather}
Like in the case of Eq.\ \eqref{eq:77} we first use the relations \eqref{eq:78} in order to transform the $H_0$ part under the last integral by partial
 integration and obtain:
\begin{gather}
  \label{eq:97}
   \{ \frac{2\varepsilon}{\hbar^2}p^2 - \frac{\varepsilon}{2}\partial^2_{\theta}\}\,V_{\psi_2\psi_1}(\theta,p) - \\
-2i\frac{\varepsilon\, mr_0^2}{\hbar}  [e^{-ip\vt/\hbar}\{\frac{ p}{m r_0^2}\,\rho_{21}(\theta;\vt) + j_{21}(\theta;\vt) \}]^{\vt = \pi}_{\vt = -\pi}, \nonumber
\end{gather} 
where
\begin{eqnarray}
  \label{eq:98}
  \rho_{21}(\theta;\vt)&=&\psi_2^{\ast}(\theta -\vt/2)\psi_1(\theta + \vt/2),\\
 j_{21}(\theta;\vt) &=& \frac{\hbar}{2i m r_0^2}\{\psi^{\ast}_2(\theta-\vt/2)\,\partial_{\theta}\psi_1(\theta +\vt/2)-~~~~~ \\
&&-(\partial_{\theta}\psi^{\ast}_2(\theta-\vt/2))\,\psi_1(\theta +\vt/2)\}. \nonumber
\end{eqnarray}
If the solutions $\psi_{1,2}(\theta)$ of Eqs.\ \eqref{eq:95} are a ``stationary'' consequence of the  time dependent equations,
\begin{equation}
  \label{eq:111}
  i\hbar \partial_t \psi_1(\theta;t) = H\psi_1(\theta;t),~~ -i\hbar \partial_t \psi_2^{\ast}(\theta;t) = H\psi_2(\theta;t),
\end{equation}
then the continuity equation
\begin{equation}
  \label{eq:112}
  \partial_t \rho_{21}(\theta;\vt;t) + \partial_{\theta}j_{21}(\theta;\vt;t) = 0
\end{equation} holds.
As $p/(mr_0^2)$ is the angular velocity $\dot{\theta}$, the term
\begin{equation}
  \label{eq:113}
  \frac{p}{m r_0^2}\rho_{21}(\theta;\vt;t)
\end{equation} in the boundary part of Eq.\ \eqref{eq:97}
represents a kind of current density, too. Thus the boundary term in Eq.\ \eqref{eq:97} represents a circular flow depending on the boundaries $\vt = \pm \pi$.

For the potential $U(\theta)$ part of $H$ under the integral of Eq.\ \eqref{eq:96} we again use the mode $U_k(\theta)$ of Eq.\ \eqref{eq:83}.
Instead of relation \eqref{eq:84} we now have
\begin{equation}
  \label{eq:103}
  U_k(\theta + \vt/2) + U_k(\theta -\vt/2) = 2\cos(k\vt/2)\,U_k(\theta).
\end{equation}
Inserting this into the potential part on the right hand side of Eq.\ \eqref{eq:96} gives the contribution
\begin{equation}
  \label{eq:99}
  2 U_k(\theta)\,\frac{1}{2\pi}\int_{-\pi}^{\pi}\frac{d\vt}{2\pi}e^{-ip\vt/\hbar}\cos(k\vt/2)\,\psi^{\ast}_2\psi_1.
\end{equation}
Using again the relation \eqref{eq:87}, the last expression can be written as
\begin{equation}
  \label{eq:100}
 2 U_k(\theta)\,\cos[(k/2)i\hbar \partial_p]\,V_{\psi_2\psi_1}(\theta,p),
\end{equation}
with
\begin{equation}
  \label{eq:101}
  \cos z = \sum_{n=0}^{\infty}(-1)^n\frac{z^{2n}}{(2n)!}.
\end{equation}
Adding the contribution \eqref{eq:100} to the term \eqref{eq:97}, inserting $\varepsilon$ from Eq.\ \eqref{eq:114} and dividing the result by $2$ finally gives
\begin{gather}
  \label{eq:102}
 \{\frac{p^2}{2mr_0^2} + U_k(\theta) - \frac{\hbar^2}{8mr_0^2}\partial^2_{\theta}\}V_{\psi_2\psi_1 }(\theta,p) + \\ \nonumber
+ \{ U_k(\theta) \sum_{n=1}^{\infty}\frac{(k\hbar/2)^{2n}}{(2n)!}\partial_p^{2n}\}V_{\psi_2\psi_1 }(\theta,p) + \\ \nonumber
+\frac{\hbar}{2 i}  [e^{-ip\vt/\hbar}\{\frac{ p}{mr_0^2}  \,\rho_{21}(\theta;\vt) + j_{21}(\theta;\vt) \}]^{\vt = \pi}_{\vt = -\pi} =
 \\= \frac{E_1+E_2}{2}V_{\psi_2 \psi_1 }(\theta,p) \nonumber
\end{gather}
For the Wigner function itself, for which $\psi_2(\vp) = \psi_1(\vp) = \psi(\vp)$ and $E_2 = E_1$, the last equation takes the form
 \begin{gather}
   \label{eq:105}
   \{\frac{p^2}{2mr_0^2} + U_k(\theta) - \frac{\hbar^2}{8mr_0^2}\partial^2_{\theta}\}V_{\psi }(\theta,p) + \\ \nonumber
+\{ U_k(\theta) \sum_{n=1}^{\infty}\frac{(k\hbar/2)^{2n}}{(2n)!}\partial_p^{2n}\}V_{\psi }(\theta,p) + \\ \nonumber
+\frac{\hbar}{2 i}  [e^{-ip\vt/\hbar}\{\frac{ p}{mr_0^2}  \,\rho(\theta;\vt) + j(\theta;\vt) \}]^{\vt = \pi}_{\vt = -\pi} 
 \\= E\,V_{\psi }(\theta,p),~~~~~~~~~~~~~~~ \nonumber 
 \end{gather} where now
\begin{eqnarray}
  \label{eq:106}
  \rho(\theta;\vt)&=&\psi^{\ast}(\theta -\vt/2)\psi(\theta + \vt/2),\\
 j(\theta;\vt) &=& \frac{\hbar}{2imr_0^2}\{\psi^{\ast}(\theta-\vt/2)\,\partial_{\theta}\psi(\theta +\vt/2)- \\
&&-(\partial_{\theta}\psi^{\ast}(\theta-\vt/2))\,\psi(\theta +\vt/2)\}. \nonumber
\end{eqnarray}
\subsection{PDEs for Wigner functions}
In the last two subsections we have derived PDEs for the Wigner function $V_{\psi}(\theta,p;t)$ or $V_{\rho}(\theta,p;t)$, Eq.\ \eqref{eq:165} for the
time evolution and Eqs.\ \eqref{eq:104} and \eqref{eq:105} for stationary systems. The question comes up immediately, whether these equation can replace
- at least in principle - the Schr\"odinger wave or Heisenberg operator equations, like path integrals in quantum mechanics or the Hamilton-Jacobi
 equation in mechanics.

 The time evolution of the Wigner function in the planar case is considered to be equivalent to that of the Schr\"odinger equation \cite{case}, mainly because both
are of first order in the time derivative and, therefore, have equivalent initial value problems. As the corresponding Eq.\ \eqref{eq:165} has the additional
boundary terms \eqref{eq:110} this problem has to be analysed anew for the cylindrical case.

As to the stationary Eqs.\ \eqref{eq:104} and \eqref{eq:105} the question arises whether they can replace the stationary Schr\"odinger equation, especially an
 eigenvalue equation. Such an eigenfunction has to obey the normalization conditions \eqref{eq:133} and \eqref{eq:27} and should have the 
properties \eqref{eq:52} -
\eqref{eq:46} for $\psi_2 = \psi_1 = \psi$. All this has yet to be analysed. In the planar case the corresponding eigenvalue problem has been positively solved
for the harmonic oscillator \cite{fair,schl}. But, of course, one has to get far beyond the harmonic oscillator in order to establish a generally attractive
 and convincing framework. There is much work ahead!

Eqs.\ \eqref{eq:165}, \eqref{eq:104} and \eqref{eq:105} have a structure which invites to attempt semi-classical approximations and classical limits: 
The left hand sides are "classsical",
i.e.\ they do not contain any $\hbar$, whereas the right hand sides contain a (formal) power series in $\hbar$ which appears to be useful for approximations.
 This was already
one of the main motives of Wigner in 1932 \cite{wig}. Many investigations followed. See, e.g., the Refs.\ \cite{vor1,ber,wig1,osb,wer,lan,schl}.
 However, one has to beware
of essential singularities in $\hbar$ \cite{vor2,hel1,hel2,hel3} which can spoil naive polynomial approximations!

\begin{acknowledgments} Special thanks go to the DESY Theory
  Group for its continuous generous hospitality since my retirement
  from the Institute for Theoretical Physics of the RWTH Aachen.
\end{acknowledgments}

\bibliography{wignerfunction}
\end{document}